\documentclass{webofc}
\usepackage[varg]{txfonts}   
\usepackage{subcaption}

\begin{document}
\title{Strangeness in astrophysics: Theoretical developments}

\author{
\firstname{Veronica} \lastname{Dexheimer}\inst{1}\fnsep\thanks{\email{vdexheim@kent.edu}} 
\and
\firstname{Krishna} \lastname{Aryal}\inst{1}\fnsep\thanks{\email{karyal@kent.edu}}
}
\institute{Department of Physics, Kent State University,Kent, OH 44242}

\abstract{%
  In this conference proceeding, we review important theoretical developments related to the production of strangeness in astrophysics. This includes its effects in supernova explosions, neutron stars, and compact-star mergers. We also discuss in detail how the presence of net strangeness affects the deconfinement to quark matter, expected to take place at large densities and/or temperatures. We conclude that a complete description of dense matter containing hyperons and strange quarks is fundamental for the understanding of modern high-energy astrophysics. 
}
\maketitle
\section{Historical Background}
\label{intro}
The history of strangeness in dense matter astrophysics started in the early 80's with a paper from Glendenning describing the role played by hyperons in neutron stars \cite{GLENDENNING1982392}. In this seminal work, it is discussed how as ``the density of matter rises, the Fermi energy of the neutrons will eventually exceed the threshold for decay, first to protons and then to the hyperons'' and that these decays are not forbidden, since astrophysical timescales are orders of magnitude larger than the one characteristic of the weak interaction.  The baryon octet is described by a modified Wallecka model and strange particle populations are shown to occupy most of the neutron star radii for massive stars, which could in this approach reach $1.8$ M$_\odot$. A few years latter, a very famous paper was written by Witten describing the possibility of stable strange quark stars \cite{PhysRevD.30.272}. In this work, the quarks are described as a free gas of massless quarks stabilized by a vacuum pressure, which is introduced to take into account the non-perturbative eﬀects of QCD. Under an adequate choice of choice bag constant, $2$ solar mass stars could easily be reproduced. In the 90's, the possibility of having kaons (and their further condensation) in neutron stars were also studied \cite{THORSSON1994693}. It was found that, it is hard to support anything larger than $1.44$ M$_\odot$ neutron stars in the presence of a large strangeness content. 

The numbers stated above immediately point out to what later became know as the hyperon puzzle \cite{Chamel:2013efa}. This relates to the fact that the softening caused by the new strange degrees of freedom turns the equation of state softer and reproduces lower neutron-star masses. Initial attempts to parametrize very stiff equations of state in order to compensate for the softening due to the strange degrees of freedom reproduced larger neutron-stars, in disagreement with laboratory and observational data. In reality, equations of state do not have to be overall soft or stiff, but can be a mixture of both by being fitted to different data at different densities. This can be constructed by tuning the strength of the attractive and repulsive interactions, and also exploring the nature of different couplings \cite{Dexheimer:2018dhb,Dexheimer:2020rlp}. In addition, it was shown that including not only hyperons, but also their respective negative parity states could still be consistent $2$ M$_\odot$ neutron stars (and other laboratory and observational data), if a crossover transition to quark matter is allowed \cite{Mukherjee:2017jzi,Motornenko:2019arp}.

\section{Different Scenarios}

The aforementioned softening of the equation of state caused by the appearance of strange hadronic and quark degrees of freedom can be seen for a few different models in Fig.~1, which was modified from Ref.~\cite{Tan:2021ahl}. Note that equations of state constructed from state of-the-art models for neutron stars generically lead to characteristic non-trivial, non-monotonic structure in the speed of sound \cite{Baym:2019iky,Guichon:1995ue,Stone:2019blq,Alford:2017qgh,Dexheimer:2008ax,Dexheimer:2009hi,Dexheimer:2020rlp,Dexheimer:2014pea}. In the particular case of Ref.~\cite{Dexheimer:2014pea}, a phase transition associated with the amount of strangeness was found to be associated with the appearance of twin stars. But is is very important to keep in mind that observational constraints for neutron stars can still be fulfilled by models that present "bumps" in this speed of sound. This is illustrated in Fig.~2, where parametric equations of state \cite{Tan:2020ics} produce mass-radius diagrams and tidal deformability-mass diagrams showing curves that are in agreement with NICER and LIGO/VIRGO data. Furthermore, bumps that present an early raise in the speed of sound (before 2 times saturation density) can generate massive stars beyond $2.5$ M$_\odot$, which points out to the possibility of the secondary object observed gravitationally in the merger GW190814 \cite{LIGOScientific:2020zkf} being a neutron star (and not a black hole). Other approaches have also been able to describe the secondary object in GW190814, in spite of its strange content (see for example Refs.~\cite{Dexheimer:2020rlp,Rather:2021azv}).

\begin{figure*}[t]
\centering
\centering
\includegraphics[width=0.5\textwidth]{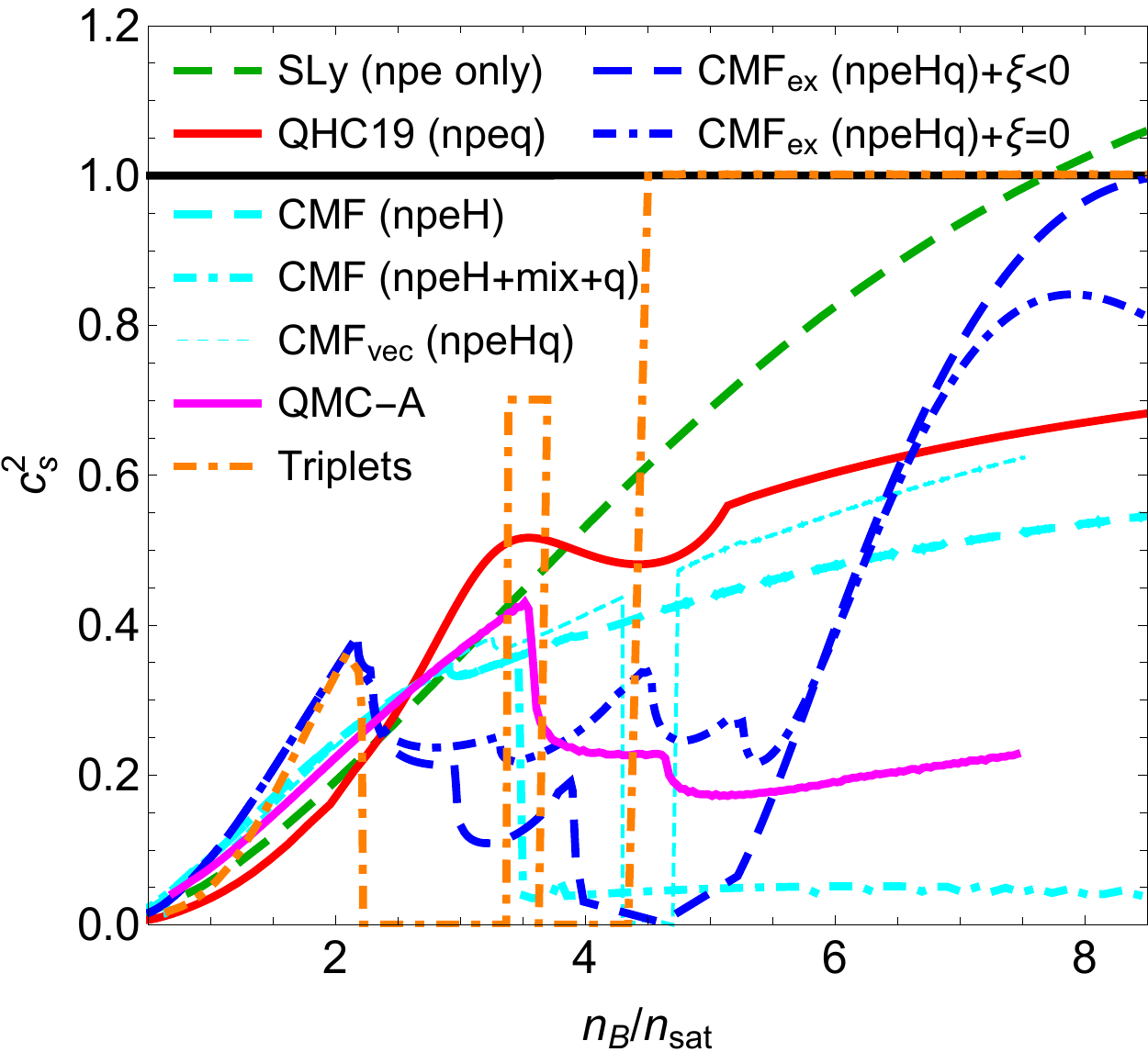}
  \caption {Speed of sound squared for the following models as a function of baryon number density in nuclear saturation units: SLy \cite{Chabanat:1997un}, QHC19
 \cite{Baym:2019iky}, CMF \cite{Dexheimer:2008ax,Dexheimer:2009hi,Dexheimer:2020rlp,Dexheimer:2014pea}, QMC-A EoS \cite{Guichon:1995ue,Stone:2019blq}, and Triplets \cite{Alford:2017qgh}. Figure modified from Ref.~\cite{Tan:2021ahl}.}
\label{EoSs}
\end{figure*}  

\begin{figure*}[t!]
\centering
\begin{tabular}{c c c}
\includegraphics[width=0.3\linewidth,trim={.05cm 0 0 0},clip]{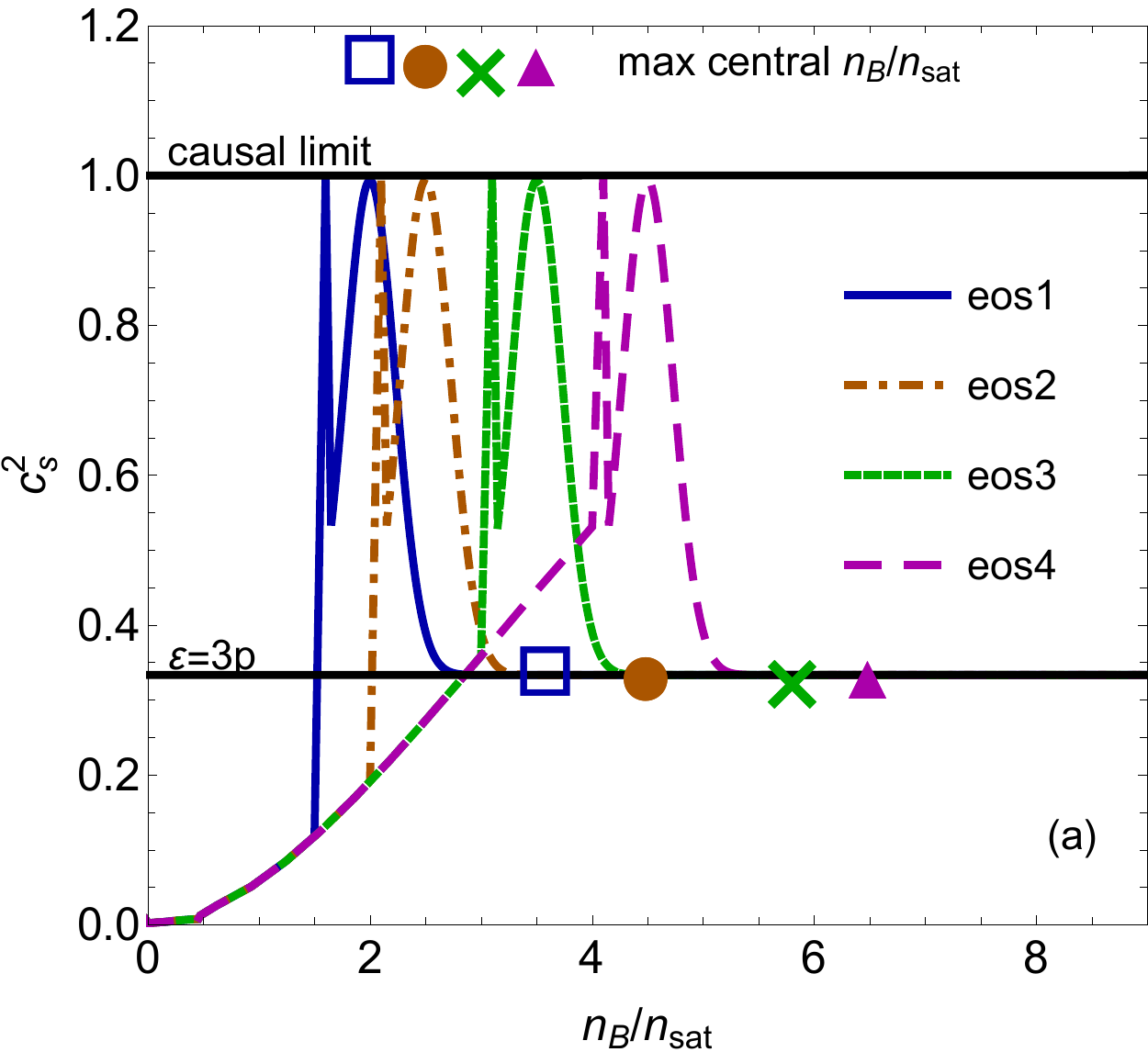} &
\includegraphics[width=0.308\linewidth]{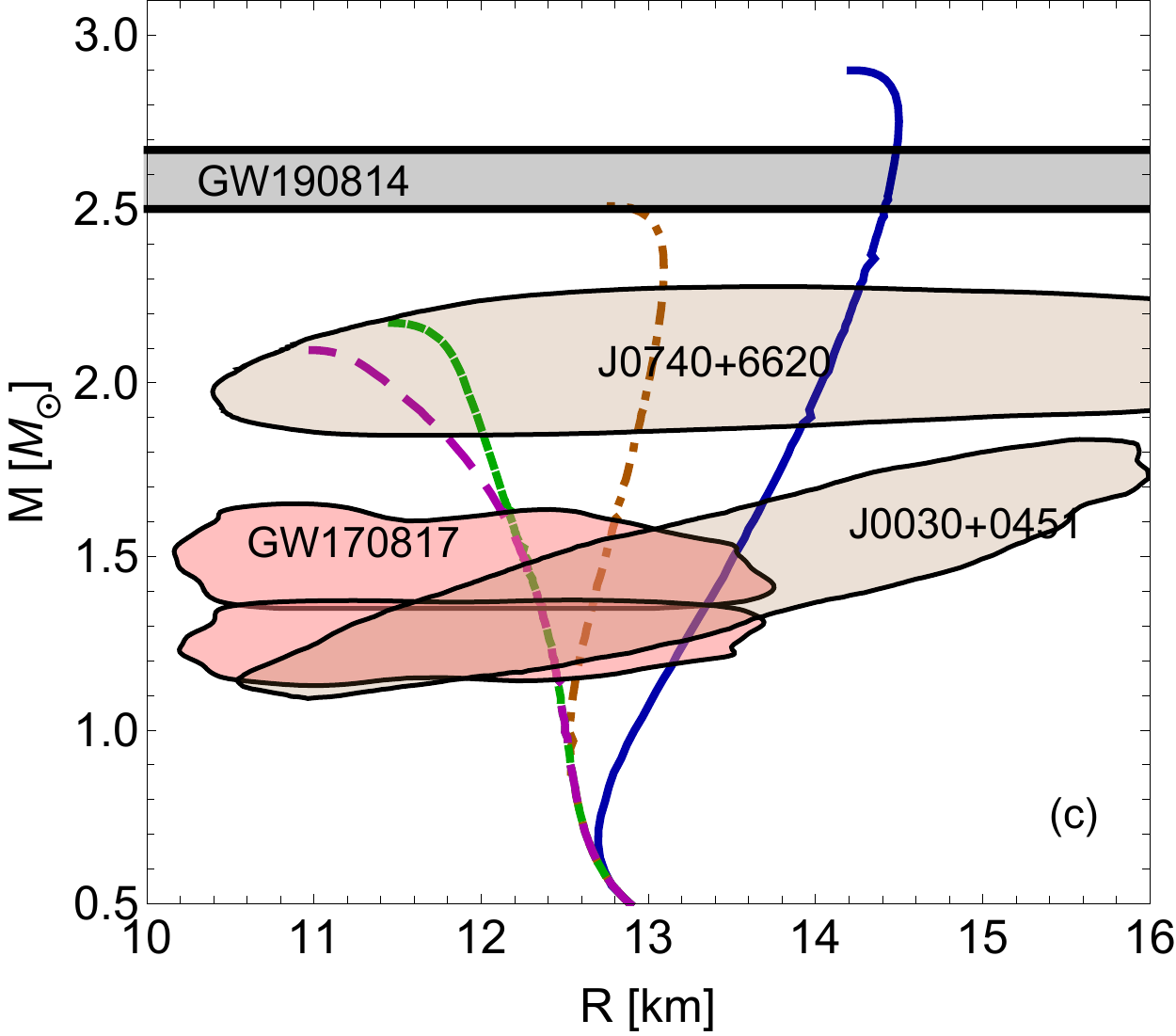} & \includegraphics[width=0.315\linewidth]{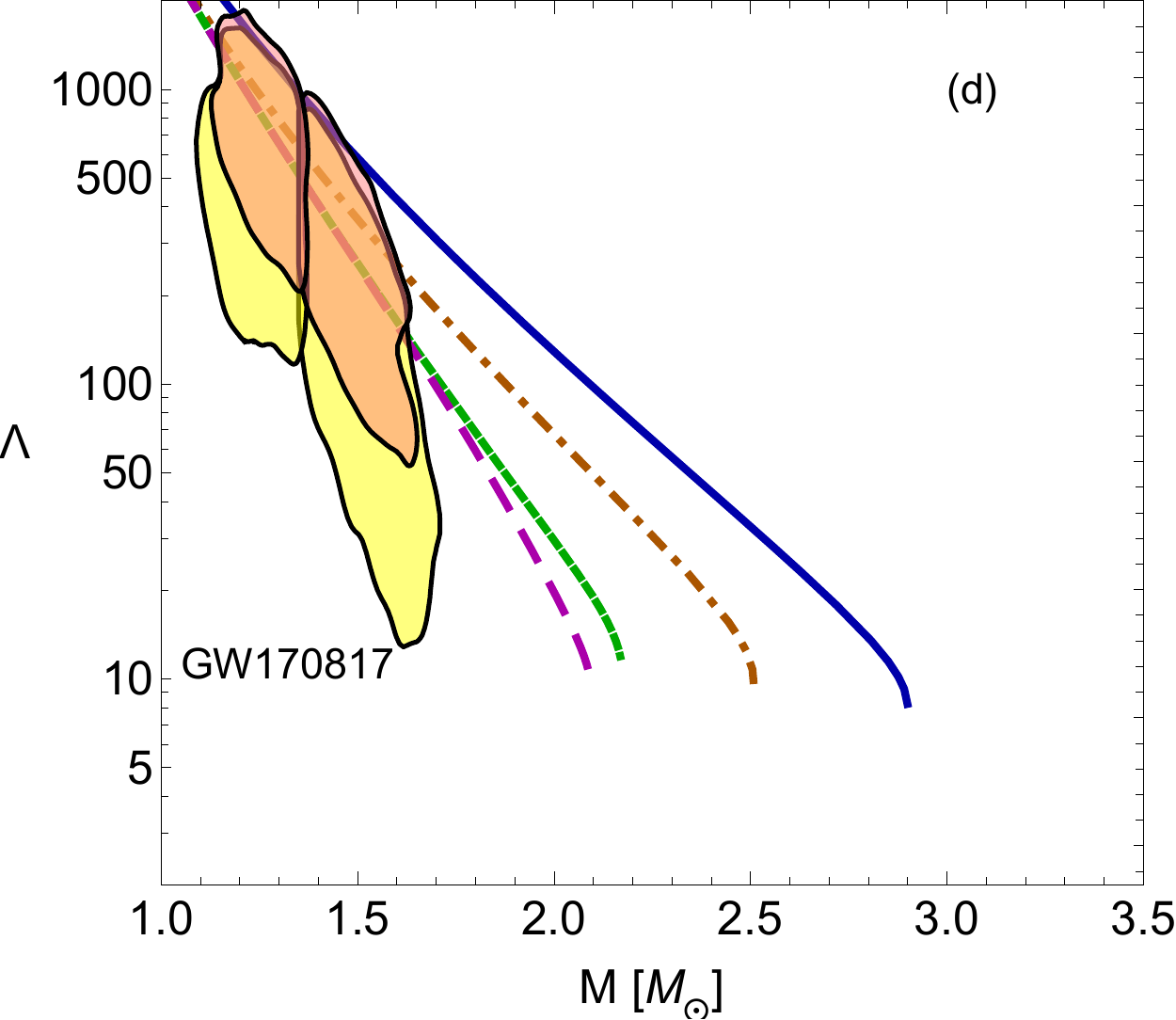}
\end{tabular}
  \caption {Speed of sound, mass-radius diagram, and tidal deformability for a family of equations of state with double peaks in the speed of sound at different locations. Symbols show the central density for the most massive star of the sequence. Figure modified from Ref.~\cite{Tan:2021ahl}}
\label{Jaki}
\end{figure*}

In supernova explosions, the presence of s-quarks were shown to turn non-explosive massive-star 3-dimensional simulations into successful ones, as strangeness affects neutrino interactions, reducing neutrino opacity, and generating higher luminosities and mean energies of neutrino emission \cite{Melson_2015}. Hyperons were also shown to affect both the density and temperature profiles inside massive exploding stars \cite{Char_2015}. In neutron-star mergers, it was shown that similar changes in neutrino luminosity are not expected to be distinguishable before high temperature regions are swallowed into the black hole, but the changes in gravitational wave frequency and amplitude before and after merger could signal hyperons \cite{PhysRevLett.107.211101}, that could
lead to a faster collapse of hypermassive stars into black holes \cite{Perego:2019adq}. The densities and temperatures found in mergers when allowing for exotic (hadronic and quark) degrees of freedom \cite{Most:2018eaw} are actually ideal conditions to generate hadronic and quark strangeness, as shown in Fig.~6 of Ref.~ \cite{Most:2019onn}.

The final scenario we describe is the matter generated in particle collisions. Although net strangeness (difference between content of strange particles and strange anti-particles) is not produced in the time frame involved in these collisions, strange and anti strange particles are produced and measured. For example, recent results from ALICE experiment in CERN give very good estimates for the $\Lambda$, $\Sigma$, and $\Xi$ hyperon potentials in medium \cite{Fabbietti:2020bfg}. Using these results, it was shown that that neutron stars with masses larger than $2.1$ M$_\odot$ can still be reproduced when the isovector coupling is rescaled \cite{PhysRevC.85.065802,PhysRevC.98.065804}.

Concerning deconfinement to (strange) quark matter, the presence of net strangeness can affect the position of the phase transition, as shown in the left panel of Fig.~3. A calculation using the Chiral Mean Field (CMF) \cite{Dexheimer:2009hi,Aryal:2020ocm,Dexheimer:2020xmh,Dexheimer:2020okt} model showed that the effect of net strangeness can affect the position of deconfinement (with respect of free energy per baryon $\widetilde{\mu}= \mu_B + Y_Q\mu_Q + Y_S\mu_S$ or baryon chemical potential $\mu_B$) by more than $40$ MeV, depending on the temperature $T$ and charge fraction $Y_Q$ (or isospin fraction $Y_I$), being larger for larger temperatures. Note that the free energy per baryon is only equal to the baryon chemical potential when either the charge fraction or charge chemical potential $\mu_Q$ and either the strangeness fraction $Y_S$ or chemical potential $\mu_S$ are zero. For zero and low temperature, the effect of strangeness is larger for lower charge (or more negative isospin fraction), corresponding to conditions achieved in chemically equilibrated neutrons stars or their mergers. Larger charge or zero/slightly positive isospin fraction are achieved in situations where beta-equilibrium hasn't taken the system away from a more energy favorable (under normal conditions of density and temperature) isospin symmetric case, corresponding to the case of particle collisions or supernova explosions. Note that the relation between charge and isospin chemical potential is not trivial (meaning always differing by exactly $0.5$) in the presence of strangeness, as seen in the deconfinement curves shown in the right panel of Fig.~3 for the zero temperature case. The relation between isospin and charge fractions depends on the strangeness fraction, which again is larger for lower charge/isospin fractions $Y_I =Y_Q - \frac{1}{2} +\frac{1}{2}Y_S$.

\subsection{Conclusions}

Give our current understanding of astrophysical environments, strangeness is not ruled out in dense matter and it is expected to appear in significant quantities already in supernova explosions, remaining present in cold catalyzed neutron stars, and generated in large amounts in hot neutron-star mergers. In this way, strangeness is expected to be important for several dynamical processes and could generate neutrinos, electromagnetic and gravitational signals expected to be measured in the near future.

Although comparisons with particle collisions could give us insight into high-energy matter, the conditions produced in this case are quite different from astrophysical scenarios, even considering that low-energy collisions overlap in temperature with neutron star mergers. These differences include the amount of charge/isospin and strangeness, both of which affect the deconfinement to quark matter. In this sense, deconfinement signals from neutron-star mergers could give us additional information about the amount of strangeness inside compact stars.

\begin{figure*}[t!]
\centering
\begin{tabular}{c c}
\includegraphics[width=0.53\textwidth,trim={.8cm 0 1.8cm 0},clip]{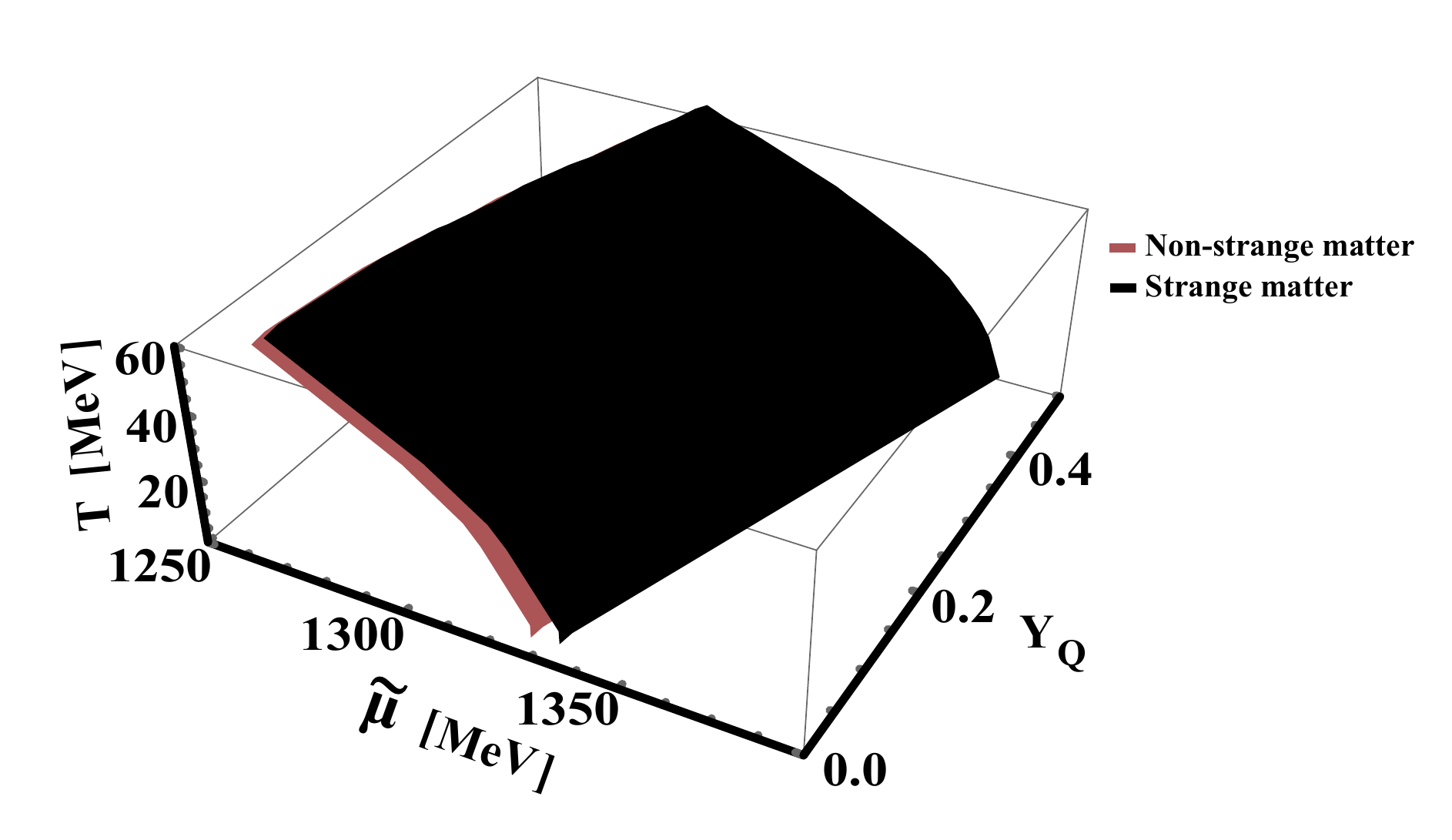} &
\includegraphics[width=0.47\textwidth,trim={.8cm 0 0 0},clip]{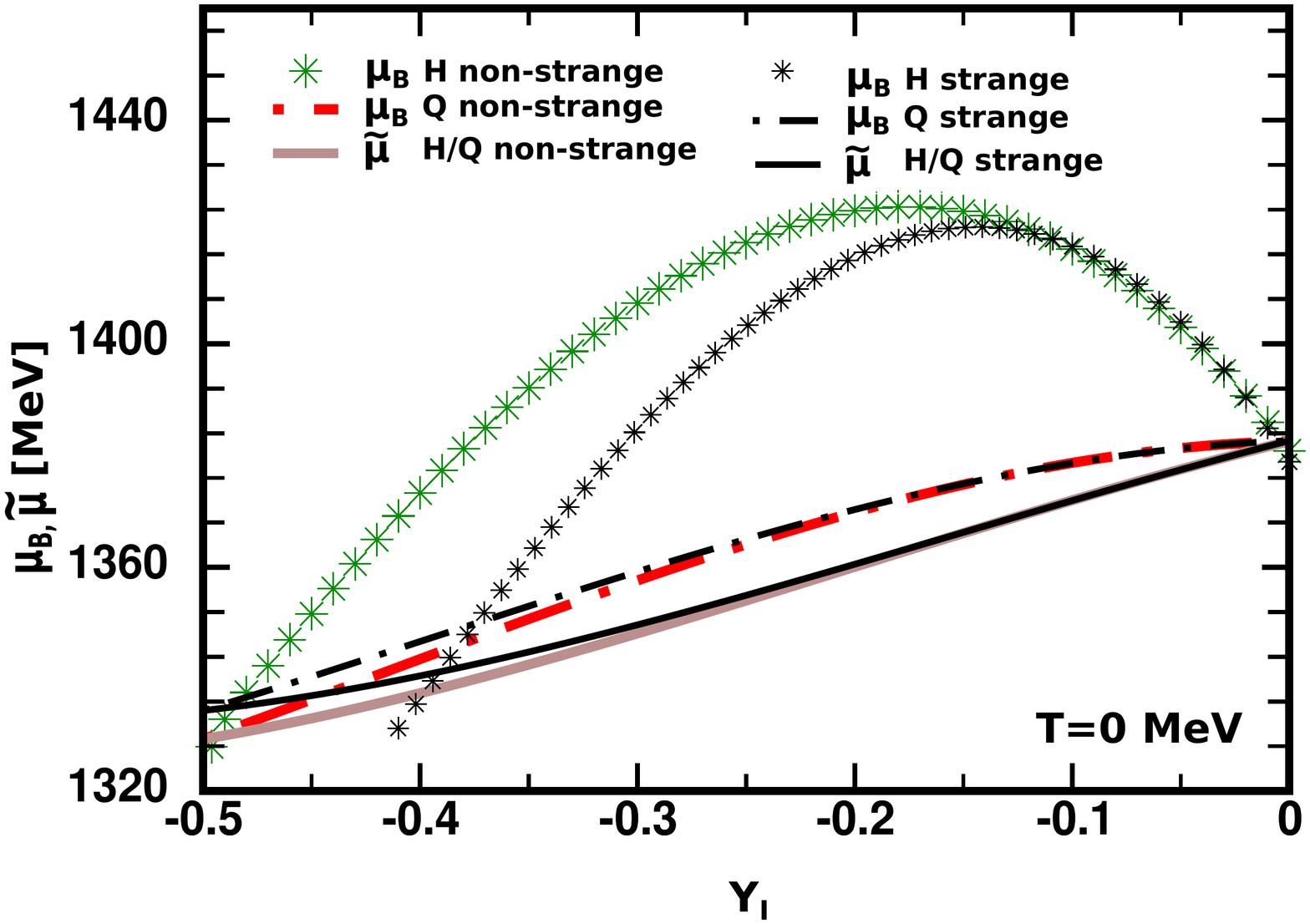}
\end{tabular}
  \caption {Phase diagrams for deconfinement to quark matter calculated from the CMF model for non-strange (color), and strange matter (black) calculated for charge fraction varying from zero to $Y_Q=0.5$. Left panel shows temperature (only until $60$ MeV), free energy per baryon $\Tilde{\mu}$, and charge fraction $Y_Q$. Right panel shows free energy per baryon $\Tilde{\mu}$, baryon chemical potential $\mu_B$ (in either side of the phase transition), and isospin fraction $Y_I$. Figure modified from Ref.~\cite{Aryal:2020ocm}.}
\label{fig2:fig}
\end{figure*} 

\section*{Acknowledgements}

Support for this research comes from the National Science Foundation under grant PHY-1748621 and PHAROS (COST Action CA16214).

\bibliography{confe}

\end{document}